\documentclass[11pt,twoside]{./atmp}

\usepackage{amsmath,amssymb, oldgerm}

\usepackage[all]{xy}
\usepackage{color}
\newtheorem{lemma}{Lemma}
\begin{document}

\title[The perturbation equation]
{The perturbation equation of a static symmetrical homogeneous space-time}

\arxurl{}

\author[Jose L. Martinez-Morales]{Jose L. Martinez-Morales}

\address{Institute De Mathematics\\
National Autonomous University De Mexico\\
A.P. 273, Administration De corers \#3\\
Cuernavaca, Morels 62251,
Mexico}  
\addressemail{martinez@matcuer.unam.mx}

\begin{abstract}
In absence of explicit solutions of the perturbation equation of a static symmetrical homogeneous space-time, the best we can do is to construct a {\it quasi-}transformation. In this framework, we solve the perturbation equation with initial data and a number of results are derived.

Far from the horizon of a black hole of even space dimension $N$, a mass-less field decays as ${r^l} {{(-{r^2}+{t^2})}^{\frac{1-N}{2}-l}}$ in space-time, where $l$ is a harmonic number of the sphere.

A relation of energy and momentum of a particle with mass in a hyper black hole is discovered and a solution to the equation of Klein-Gordon in the metric of Schwarzschild-Tangherlini with initial data on the hypersphere is proposed. Also, the Green's function of the Klein-Gordon equation in Schwarzschild coordinates is calculated. This function is a sum on the harmonic modes of the sphere. The first term is a double integration on the spectrum of energy and the momentum of the particle. Far from the horizon, the double integration is approximated by an integration on a line defined by the relation of energy and momentum of a free particle. From here, the potential of Yukawa is derived. Finally, the linear perturbation equations are derived and solved exactly.

The master equation with initial data of a small perturbation in a static symmetrical homogeneous space-time, like a (possibly higher dimensional) Schwarzschild black hole, is studied. A main statement of the article is that for each harmonic mode of the horizon there are two solutions that behave similarly at large. In the basic mode, the asymptote of a field (an eigentensor of the Lichnerowicz operator, for example) decays at infinity according to the universal law ${{(-{r^2}+{t^2})}^{\frac{1-N}{2}}}$. These solutions occur in an integral form.

The analysis we present is of a small perturbation to the full static symmetrical solution. The higher order perturbations will appear in a sequel. We determine independently perturbations of the space-time in dimension 1+$N\ge$ 4, where the system of equations can be reduced to a master equation --- a tensor differential equation. The solutions are integral transformations which in some cases reduce to explicit functions.
\end{abstract}

\maketitle
\section{Introduction}
The manuscript deals with the evolution of fields in background space-times with certain symmetry, mainly the higher-dimensional Schwarzschild black holes (or general spherically symmetric curved) geometries in a linearized approach. We solve the Klein-Gordon equation and the equation for metric perturbation introducing a phase space. A number of results are derived.

Since it is an important problem, some motivations are described below. The reader should probably like to read the original works by Reggae and Wheeler and others, for instance: General theory of perturbation analysis and motivation \cite{Regge} \cite{Vishveshwara} \cite{ZerilliD} \cite{Davis}, Quasi-normal modes \cite{Kokkotas} \cite{Lemos}, and Late-time Tails in black hole backgrounds \cite{Price} \cite{Leaver} \cite{Lett} \cite{D} \cite{Yoshida}.

Space-time is described by
\begin{equation}\label{1} {\hbox{\rm Ric}_{\hbox{\textgoth g}}}=\frac R4\hbox{\textgoth g},
\end{equation} where $\hbox{\textgoth g}$, $R$ and $\hbox{\rm Ric}_{\hbox{\textgoth g}}$ are the metric, Ric scalar and Ric tensor of empty universe. The Einstein equations (\ref{1}) can be derived from the Einstein-Hilbert action
\begin{equation}
\frac{1}{16\pi}\int d^4 x ~\sqrt{-g}R
\end{equation}
where $g=\det \hbox{\textgoth g}$.

A consistent theory of quantum gravity, such as string theory, seems to require the existence of higher dimensions, which in order to not contradicting observational evidence must be compact on small scales. String theory has made some important progress in explaining the entropy of certain black holes by counting microscopic degrees of freedom, and thus scenarios with extra, compact dimensions must be taken seriously. In turn, some of the most interesting objects to be studied within string theory are those that possess event horizons, possibly extended in the compact extra dimensions.

Higher-dimensional theories of gravitation have recently attracted much attention, and static symmetrical homogeneous solutions have played an important role in revealing various aspects of gravitation peculiar to higher dimensional systems \cite{Myers} \cite{Witten} \cite{Dimopoulos} \cite{Giddings}. A noteworthy fact is that, compared to the severely constrained situation in four space-time dimensions \cite{Israel} \cite{Carter} \cite{Robinson}, higher dimensional black holes can have many very different varieties \cite{GibbonsLett} \cite{GibbonsD} \cite{Emparan} \cite{GibbonsHartnoll} \cite{GibbonsHartnollPope}. For this reason, it is important to study the dynamics of fields \cite{Cardoso} \cite{Ida} \cite{HorowitzHubeny} \cite{Konoplya} \cite{Molina} and space-time perturbations \cite{GibbonsHartnoll} \cite{GibbonsHartnollPope} \cite{Hartnoll} in such black hole backgrounds and to find stable solutions. Amongst the various types of static symmetrical homogeneous solutions, a natural higher dimensional generalization of a black hole --- also known as a Schwarzschild-Tangherlini black hole \cite{Tangherlini} --- has been assumed to be (or believed to be) stable, like its four-dimensional counterpart, and for this reason it has been conjectured that this black hole describes a final equilibrium state of evolution of an isolated system, or a certain class of unstable static solutions in higher dimensions. However, the question of the asymptote of a perturbation of a black hole itself is not a trivial matter, and, to our knowledge, it has not yet been fully analyzed in higher dimensions.

Our manifold will be the mathematical manifold $M^{1+N}$, the first layer of the would-be physical space-time of general relativity. The space-time is assumed to be $1+N$ dimensional. It has been common practice in recent years to denote $n$ as the number of extra dimensions, besides the usual four. We are consistent with Ref. \cite{Myers}. In the mathematical literature about topological spaces, it is always implicitly assumed that the entities of the set can be distinguished and considered separately (provided the Hausdorff conditions are satisfied), otherwise one could not even talk about point mappings or homeomorphisms. It is well known, however, that the points of a homogeneous space cannot have any intrinsic individuality\footnote{As Hermann Weyl \cite{11a} puts it: ``There is no distinguishing objective property by which one could tell apart one point from all others in a homogeneous space: at this level, fixation of a point is only by a {\it demonstrative act} as indicated by terms like {\it this} and {\it there}.''}.

For concreteness, we consider as background the Cartesian product of a two-dimensional orbit space ${\mathcal N}\sp2$ and a horizon $M\sp{N-1}$ with $N-$1 compact dimensions, where 1+$N$ is the total space-time dimension and $N\ge$ 4 (in order to avoid space-times with one or less compact spatial dimensions where the presence of a massive source is inconsistent with asymptotic flatness).

Tangherlini's idea was to consider pure gravity in 1+$N$ dimensions. The (1+$N$)-dimensional action is simply
\begin{equation}
\int d^{1+N}x\sqrt{-g} R
\end{equation}
where now $\hbox{\textgoth g}$ is the 1+$N$ dimensional metric, and $R$ is the corresponding Ric scalar. We have the original first two dimensions labeled with coordinates $x^{\mu}$ where $\mu=0,1$. The $N-$1 remaining dimensions are compacted on a hyper sphere and are labeled by the coordinate $\chi\in S\sp{N-1}$.

Now we can expand the metric as a spherical harmonics series of the form
\begin{equation}
 \hbox{\textgoth g}(x, \chi)=\sum_{l=0}\sp\infty\sum_{m_l} \hbox{\textgoth g}^{({m_l})}(x)H_{m_l}(\chi).
\end{equation}
We find that we get an infinite number of fields in the first two dimensions. Modes with ${m_l}\neq 0$ correspond to massive fields with mass $|{m_l}|/\hbox{vol}S\sp{N-1}$. The zero mode corresponds to a mass-less field. As we take $\hbox{vol}S\sp{N-1}$ to be smaller and smaller we see that the mass of the first massive field becomes very large. This means that if we compact on a small enough hyper sphere we can truncate to mass-less modes in the two-dimensional theory. We can only see the extra dimensions by exciting massive modes which are at energies beyond our reach.

Let us now focus on the zero mode $\hbox{\textgoth g}\sp{(0)}(x)$, the metric of ${\mathcal N}\sp2$. We could define $\hbox{\textgoth g}$ so that {\textgoth g}$\sp{(0)}$ is the two-dimensional field in the Schwarzschild- Tangherlini metric. In order that our results are more transparent we introduce some notation. We denote by $r$ the ``radial" coordinate, this is the coordinate away from the horizon. In the vicinity of the static homogeneous horizon it is useful to introduce ``spherical" coordinates $\chi$ as well. We will actually introduce the components of the metric in the following way:
\begin{equation}
\label{2}
 \hbox{\textgoth g}\sp{(0)}=-g_{00}(r)d t\otimes d t+g_{11}(r)d r\otimes d r,\qquad \hbox{\textgoth g}=\hbox{\textgoth g}\sp{(0)}+r\sp2\hbox{\textgoth h},
\end{equation}
where $g_ {00} $ and $g_ {11} $ are two derivable functions of the variable $r$, and
\begin{equation}
\hbox{\textgoth h}=h(z)_{i j}dz\sp id z\sp j
\end{equation}
denotes the metric of the horizon\footnote{Accordingly, we also refer to the metric (\ref{2}) as a {\it maximally symmetric static homogeneous space-time} \cite{GibbonsHartnoll} \cite{GibbonsHartnollPope} \cite{Birmingham}. Ishibashi and Kodama briefly comment on such a generalized static homogeneous space-time in section 4 of \cite{Ishibashi}. }, i. e., it is the metric of the Einstein space ($M\sp{N-1}$, {\textgoth h}) whose Ric curvature is
\begin{equation}
\hbox{\rm Ric}_{\hbox{\textgoth h}}=\frac{S}{N-1}\hbox{\textgoth h},
\end{equation}
$S$ being the Ric scalar of the horizon.

We have truncated to the mass-less fields. Tangherlini type compactifications can be more complicated than simply compacting on a hyper sphere. The important thing is that the extra dimensions are small so that we do not excite massive modes. We can truncate to mass-less modes and read off the effective theory in the first two dimensions.

We need not restrict ourselves to just two dimension either. In fact, higher dimensions have become very fashionable in the last twenty years, mainly due to the success of {\it string theory} as a possible quantum theory of gravity. At the quantum level, boson string theory is only consistent\footnote{Actually, boson string theory contains a tachyon, but we will ignore that here.} in twenty-six (!) dimensions, although this figure is reduced to ten when we introduce super symmetry. Furthermore, there are five distinct string theories which can be viewed as different elements of an embracing new theory, M-theory~\cite{Witten:Mtheory} \cite{Schwarz:Mtheory} \cite{Duff:Mtheory}. M-theory lives in eleven dimensions and has eleven-dimensional super gravity as its low energy limit.

Traditionally we achieve the reduction down to the first two dimensions using Tangherlini techniques. If we start with a $(1+N)$-dimensional theory, we compact on a small ($N-1$)-dimensional manifold. Different manifolds generally give different effective theories in the first two dimensions. The one thing all of these manifolds have in common is that they are very small, and compact.

There is, however, an alternative to Tangherlini compactification. This is the idea that we live on something called a {\it braneworld}, where the extra dimensions can be infinite.

In 1+$N-p$ space-time dimensions, an event horizon can be topologically a sphere $S^{N-1-p}$, but when extended to $p$ extra dimensions, it could naturally have topologies either $S^{N-1}$ in which case it is a higher dimensional black hole, or $S^{N-1-p}\times R ^{\,p}$ being then a black $p$-Brana or a black string in the $p$=1 case \cite{Myers} \cite{Tangherlini} \cite{Horowitz}. The $R^{\,p}$ topology of the transverse dimensions can be compacted giving flat toroidal topological spaces $T^{\,p}$.

One of the first steps towards understanding these extended higher dimensional solutions is to investigate their classical stability against small perturbations. If a solution is unstable, then it most certainly will not be found in nature (unless the instability is secular) and the solution looses most of its power. Of course the next question one must ask is what is the final stable result of such instability. Now, in (1+$N-p$)-space-time dimensions, the Schwarzschild-Tangherlini geometry is stable against all kinds of perturbations, massive or mass-less \cite{Cardoso} \cite{Regge} \cite{KodamaA} \cite{Ishibashi} \cite{KodamaProg} \cite{124017} \cite{024018} \cite{Berti}. On the other hand, quite surprisingly, Gregory and Laflamme \cite{Gregory} showed that this is not the case for higher dimensional black branes and black strings. These objects are unstable. The kind of black $p$-branes originally studied in \cite{Gregory} were solutions of ten dimensional low energy string theory with metric of the form
\begin{equation}
d s^2=d s^2_{\rm Schwa}+d x^id x_i,
\end{equation}
where $d s^2_{\rm Schwa}$ stands for the (1+$N-p$)-dimensional Schwarzschild- Tangherlini line element, the $x^i$ are the coordinates of the compact dimensions, and $i$ runs from $1$ to $p$. The total dimension of the space-time is 1+$N$. In string or super gravity theories one takes 1+$N=10$ or 1+$N=11$, but for generality one can leave it as a free natural parameter.

The specific metric we shall be interested in is given by a time-independent symmetric metric that can be written in Schwarzschild-Tangherlini coordinates \cite{Wang} as (\ref{2}).

After this general explanation of our background space-times, we give in the next section our formalism for gravitational perturbations, and the basic strategy of our analysis of the asymptote of a perturbation. At this stage, our arguments are not limited to the case of black holes, but are equally applicable to any case of maximally symmetric static homogeneous space-times with a non-vanishing cosmological constant.
\section{Solution of the perturbation equation of a\\
static symmetrical homogeneous space-time}
\label{Seccion 2}
The study of symmetric perturbations of the Schwarzschild-like solution is reduced to solving a single partial differential equation (\ref{3}). We attempt to solve this perturbation equation with initial data in a general static symmetrical homogeneous space-time.

Deriving the analytic formula for the solutions is important and useful for further study. The computation of gravitational radiation was done only numerically in a preceding study, e.g. Ref \cite{Berti}.

The first decision to be made is to choose our coordinates and ansatz for the analysis. We choose to work with one coordinate patch for the whole space-time. In our coordinates (where the metric in the $r$ line is in the form $g_{11}(r)$) the coordinate size of the horizon is an invariant and hence the coordinate patch does not change.

Consider a function $f$ of the variable $r$.

We denote by $\Delta_{\hbox{\textgoth h}}$ the Laplace operator in the metric {\textgoth h}.

Consider a base $ \{E_e \} $ of eigen-vectors of $\Delta_{\hbox{\textgoth h}}$.

Let us consider a field with configuration space $M\sp{1+N}$ with $N$ even and global coordinates $\phi$($r$, $t$, $\chi$) described by a singular Euler-Lagrange equation,
\begin{equation}
\label{3}
\left(\frac{1}{{g_{00}}(r)}\frac{\partial\sp2}{\partial t\sp2}-\frac{{r^{1-N}}}{{\sqrt{{g_{11}}(r)}} {\sqrt{{g_{00}}(r)}}}\frac\partial{\partial r}\left(\frac{{r^{N-1}} {\sqrt{{g_{00}}(r)}}}{{\sqrt{{g_{11}}(r)}}}\frac\partial{\partial r}\right)+\frac{\Delta_{\hbox{\textgoth h}}}{{r^2}}+f(r)\right)\phi=
\end{equation}
\[
=0.
\]
The symmetry of the background guarantees that time and spatial variables will separate. In other words, if we rewrite equation (\ref{3}) in $r$ and $t$ coordinates we find that the coefficients of the differential equation do not depend on time $t$, and so the problem can be reduced to one dimension by applying a formal transform with respect to the time variable $t$.

However, the perturbation equation (\ref{3}) is not invariant under simultaneous rescaling of time and space variables $r\mapsto r k$, $t\mapsto kt$. This lack of invariance reflects the presence of mass/length scale in the model. The solution is most conveniently analyzed in terms of the phase space coordinates $k$, $\omega$, $e$.

In absence of explicit solutions of the perturbation equation, the best we can do is to construct a solution by performing a {\it quasi-}transformation from coordinate variables ($t$, $r$, $\chi$) to phase space variables ($\omega$, $k$, $e$). On the other hand, such transformation must have the remarkable property that, in the {\it special regimes} with $\omega(r, t, \chi ) \approx k$, the variables $(r, t, \chi)$ and $(k, \omega, e)$ form a basis of off-shell observables for the gravitational field {\it even if} the solution of the perturbation equation is not known.
\paragraph{Basic set-up.}\rm We denote by $B_o$ the function of Bessel of order $o$, where
\begin{equation}
o={\sqrt{e+{\left(-1+\frac N2\right)}^2}}.
\end{equation}
If $ \{E_e \} $ is a discreet set (continuous set), then $ \sum_e$ denotes a sum (an integral).

For the transform of $\phi$ we have
\begin{equation}
F(k, \omega, e)={\mathcal T}\phi(r, t, \chi),
\end{equation}
with the inverse transformation being
\begin{equation}
\label{4}
{r^{1-\frac{N}{2}}}\sum_e E_e\int_{-\infty}\sp\infty\exp\Big({\sqrt{-1}}t \omega \Big)\int_0\sp\infty {B_o}(r k) F(e,k,\omega )d k d\omega.
\end{equation}
That is to say, we propose a solution of the equation (\ref{3}) of the form (\ref{4}) so the derivatives of the original problem will enter as multiplicative terms.

A vector $E_e$ satisfies
\begin{equation}
\label{5}
\Delta_{\hbox{\textgoth h}}E_e=e E_e.
\end{equation}
The functions of Bessel satisfy
\begin{eqnarray}
\label{6}
2B_o'&=&B_{o-1}-B_{1+o},\\
\label{7}
\frac{2o}x B_o(x)&=&B_{o-1}(x)+B_{1+o}(x).
\end{eqnarray}
We substitute (\ref{4}) into (\ref{3}).

We use (\ref{5}), (\ref{6}) and (\ref{7}).

(\ref{3}) becomes
\[
{r^{-N/2}}\sum_e E_e\int_{-\infty}\sp\infty\exp\Big({\sqrt{-1}}t \omega \Big)\int_0\sp\infty F(e,k,\omega )
\]
\[
\Bigg(\frac{r k B_o'(x)|_{x=r k} \frac d{d r} {{\log\sqrt{\frac{{g_{11}}(r)}{{g_{00}}(r)}}}}}{{g_{11}}(r)}+{B_o}(r k) \Bigg(\frac{e (1-\frac{1}{{g_{11}}(r)})}{r}+r \bigg(f(r)+\frac{{k^2}}{{g_{11}}(r)}
\]
\begin{equation}
\label{8}
-\frac{{{\omega }^2}}{{g_{00}}(r)}\bigg)+\frac{\frac d{d r} {{\log(\frac{{g_{00}}(r)}{{g_{11}}(r)})}^{\frac{1}{2}(-1+\frac{N}{2})}}}{{g_{11}}(r)}\Bigg)\Bigg)d k d\omega=0.
\end{equation}
We suppose that for each $e$ and each $ \omega$,
\begin{equation}
\lim_{k\to0}k F(e,k,\omega ) B_o'(x)|_{x=r k}=\lim_{k\to\infty}k F(e,k,\omega ) B_o'(x)|_{x=r k}.
\end{equation}
We integrate by parts $ \int_0 \sp \infty k F (e, k, \omega) B_o'(x)|_ {x=r k} d k$.

(\ref{8}) becomes
\[
{r^{-N/2}}\sum_e E_e\int_{-\infty}\sp\infty\exp\Big({\sqrt{-1}}t \omega \Big)\int_0\sp\infty {B_o}(r k)
\]
\[
\Bigg(F(e,k,\omega ) \Bigg(\frac{e (1-\frac{1}{{g_{11}}(r)})}{r}+r \bigg(f(r)+\frac{{k^2}}{{g_{11}}(r)}-\frac{{{\omega }^2}}{{g_{00}}(r)}\bigg)+\frac{\frac d{d r} {{\log(\frac{{g_{11}}(r)}{{g_{00}}(r)})}^{\frac{N}{4}}}}{{g_{11}}(r)}\Bigg)
\]
\begin{equation}
+\frac{k \frac d{d r} {{\log\sqrt{\frac{{g_{11}}(r)}{{g_{00}}(r)}}}} \frac\partial{\partial k}F(e,k,\omega )}{{g_{11}}(r)}\Bigg)d k d\omega=0.
\end{equation}
Therefore,
\[
\Bigg(F(e,k,\omega ) \Bigg(\frac{e (1-\frac{1}{{g_{11}}(r)})}{r}+r \bigg(f(r)+\frac{{k^2}}{{g_{11}}(r)}-\frac{{{\omega }^2}}{{g_{00}}(r)}\bigg)+\frac{\frac d{d r} {{\log(\frac{{g_{11}}(r)}{{g_{00}}(r)})}^{\frac{N}{4}}}}{{g_{11}}(r)}\Bigg)
\]
\begin{equation}
\label{9}
+\frac{k \frac d{d r} {{\log\sqrt{\frac{{g_{11}}(r)}{{g_{00}}(r)}}}} \frac\partial{\partial k}F(e,k,\omega )}{{g_{11}}(r)}\Bigg)=0.
\end{equation}
We define
\begin{equation}
\label{10}
\omega(k)=\surd \Bigg(\Bigg(f(r)+\frac{e (1-\frac{1}{{g_{11}}(r)})}{{r^2}}+\frac{{k^2}}{{g_{11}}(r)}\Bigg) {g_{00}}(r)+\frac{(\frac N2+o) \frac d{d r} \frac{{g_{00}}(r) }{{g_{11}}(r)}}{2 r}\Bigg).
\end{equation}
Then Equation \ref {9} is as
\begin{equation}
\label{11}
\Bigg(F(e,k,\omega ) \Bigg(\frac{{{g_{11}}(r)}({\omega }^2-\omega(k)^2)}{{r\frac d{d r} {{\log\sqrt{\frac{{g_{11}}(r)}{{g_{00}}(r)}}}}}}+o\Bigg)-k\frac\partial{\partial k}F(e,k,\omega )\Bigg)=0.
\end{equation}
Assume the support of the function $ F $($ e $, $ \omega $, $ k $) lies in a narrow neighborhood of the curve
\[
\omega^2=\omega(k)^2.
\]
Then Equation \ref {11} is approximately equal to
\[
o F(e, \omega, k)-k \frac\partial{\partial k}F(e, \omega, k)=0.
\]
Therefore, there exists a function $ G $ such that
\[
F(e, \omega, k)={k^o} G(e, \omega).
\]
Therefore, (\ref {4}) is as
\begin{equation}
\label{12}
{r^{1-\frac{N}{2}}}\sum_e E_e\int_{-\infty}\sp\infty\exp\Big({\sqrt{-1}}t \omega \Big)G(e, \omega)\int_0\sp\infty {k^o}{B_o}(r k)  d k d\omega.
\end{equation}
Set
\[
t=0.
\]	
Then (\ref {12}) is as
\begin{equation}
\label{13}
{r^{1-\frac{N}{2}}}\sum_e E_e\int_{-\infty}\sp\infty G(e, \omega)\int_0\sp\infty {k^o}{B_o}(r k)  d k d\omega.
\end{equation}
Then (\ref {12}) solves Equation \ref {3} with initial condition (\ref {13}).
\subsection{Extinction of the field at infinity}
Now, the usefulness of this approach with respect to previous ones is that we can derive the late-time behavior of fields in this background (see Ref. \cite{Yoshida}) with our formalism. The reader must be aware of the relevant literature on the subject, and of what has been done in recent years.

Suppose that
\begin{equation}
\lim_{r\to\infty}f(r)=0\quad\hbox{and}\quad\lim_{r\to\infty}g_{**}=1,\quad*=0, 1.
\end{equation}
Then
\begin{equation}
\lim_{r\to\infty}\omega(k)=k.
\end{equation}
We calculate
\begin{equation}
\int_0\sp\infty\exp\Big({\sqrt{-1}}t k \Big) {k^o}{B_o}(r k) d k=-\frac{{{(-2 r)}^o} {{(-{r^2}+{t^2})}^{-\frac{1}{2}-o}} \Gamma(\frac{1}{2}+o)}{{\sqrt{-\pi }}}.
\end{equation}
The previous discussion applies to a class of globally hyperbolic, topologically trivial, non-compact (asymptotically flat at spatial infinity) space-times of the type of Christodoulou-Klainermann ones \cite{20a}. In them we have \cite{19a} \cite{19b} \cite{19c} \cite{8a}:

1) The imposition of suitable boundary conditions on the fields and the gauge transformations of metric gravity reduces the asymptotic symmetries at spatial infinity to the asymptotic Poincare group. The asymptotic implementation of Poincare group makes the general-relativistic definition of angular momentum and the matching of general relativity with particle physics.

2) The boundary conditions of point 1) require that the leaves of the foliations associated with the admissible 1+$N$ splittings of space-time must tend to Minkowski space-like hyper-planes asymptotically orthogonal to the (1+$N$)-momentum in a direction-independent way. This property is concretely enforced by using a technique introduced by \cite{16a} \cite{16b} for the selection of space-times admitting asymptotically flat (1+$N$)-coordinates at spatial infinity.

These facts entail that {\it there is an effective evolution in the mathematical time} $t$ which parametrizes the leaves of the foliation associated with any 1+$N$ splitting. 

The idea behind a new quantization scheme is to arrive directly to the physical Hilbert space by quantizing only the observables of the system and treating the configuration variables as {\it c-numbers} (like {\it time} in the time-dependent Schrodinger equation; the momenta become derivatives with respect to the configuration variables, like the energy is replaced by the time derivative). In gravity this scheme would make sense only if the configuration variables are coordinate-independent. There will be as many coupled Schrodinger equations as configuration variables (gauge invariant) and the wave function will depend on as many {\it times} (besides the standard one) as configuration variables. Every line in this parameter space will correspond to a configuration of the classical theory.
\section{Decay of a mass-less field far from the horizon of a black hole}
Now consider the evolution of a mass-less scalar field $\phi$ in the background described by (\ref{2}). The evolution is governed by the curved space Klein-Gordon equation
\begin{equation}
\label{14}
\frac1{\sqrt{-g}}\partial_\alpha\sqrt{-g}g\sp{\alpha\beta} \partial_\beta \phi= 0.
\end{equation}
The metric appearing in (\ref{14}) should describe the geometry referring to both the black Brana and the scalar field, but if we consider that the amplitude of $\phi$ is so small that its contribution to the energy content can be neglected, then the metric (\ref{2}) should be a good approximation to $g_{\alpha \beta}$ in (\ref{14}). We shall thus work in this perturbation approach. It turns out that it is possible to simplify considerably equation (\ref{14}) if we separate the horizon variables from the radial and the time variables, as is done in four dimensions \cite{Brill} \cite{Teukolsky}. For higher dimensions we follow \cite{Ida}.

The evolution of a minimally coupled scalar field $\phi$ is described by the mass-less Klein-Gordon equation (\ref{8}) \cite{Gan} \cite{Cao} \cite{Wazwaz} \cite{Tian} \cite{Chen}, with
\[
f=0,
\]
where (\ref{8}) is a partial differential operator that contains information about the initial shape of the wave packet at $t$=0. The explicit form of the operator (\ref{8}) is the simplest in this general setting.
 
In Section \ref{Seccion 2}, the linear perturbation equation is derived and solved.

Setting
\[
e=l(l+N-2),
\]
we see that far from the horizon a mass-less field decays as ${r^l} {{(-{r^2}+{t^2})}^{\frac{1-N}{2}-l}}$ in time, where $l$ is the harmonic mode of the sphere. This indicates that it behaves like $\sim t^{1-N-2l}$ for fixed $r$.  Ref. \cite{Yoshida} clarify that the field behaves like $\sim t^{1-N-2l}$ for even $N$, and $\sim t^{5-3N-2l}$ for odd $N$.

The situation for the Tangherlini hyper sphere is clear. Near the hyper sphere the fluctuations in the field behave in the same way as for flat hyper spheres. However, the field does vanish at the point at infinity, and close to it the field perturbations start to decrease. Since this point lies far from the hyper sphere we might yet believe that gravity is localized at low enough energies. At finite temperature we could hide the point at infinity behind another black hole horizon.
\section{The master equation of a homogeneous space-time}
Why a fluctuation about a black hole is regular at infinity? This is an important problem in gravitation and has been studied for long time. Although we do not have a definite answer to this question yet, we have rather well established answers to the questions of stable (unstable) Gaussian fluctuations about our black hole, which are summarized as the instabilities of gravitation.

In local particle physics problems in the present universe, the assumption of the asymptotic flatness will be a good approximation. However, the success of the inflationary universe model implies that the cosmological constant cannot be neglected even locally in the early universe. Further, recent developments in unifying theories suggest the possibility that the universe has dimensions higher than four on microscopic scales \cite{Randall} \cite{Antoniadis} and that mini black holes might be produced in elementary particle processes in colliders as well as in high energy cosmic shower events. Therefore, it is a quite important problem whether the regular character of a fluctuation about a black hole also holds in the non-vanishing cosmological constant case and/or in higher dimensions.

Here we explain the derivation of the master equation for the perturbation of a higher-dimensional Schwarzschild black hole. Our purpose is to study and present an analytic procedure to obtain solutions for the master equation of a small perturbation in a static symmetrical homogeneous space-time. We investigate the asymptote of a linear perturbation of higher dimensional space-times in the framework of a gauge-invariant formalism for gravitational perturbations of maximally symmetric static homogeneous space-times. This formalism was recently developed by Ishibashi and Kodama \cite{Ishibashi}.

In the four-dimensional case, the perturbation analysis of a black hole was first carried out by Reggae and Wheeler \cite{Regge}, and its stability was essentially confirmed by Vishveshwara \cite{Vishveshwara}, followed by further detailed studies \cite{Price} \cite{20} \cite{21}. The basic observation of those works is that unstable perturbations, whose growth in time is unbounded, are initially divergent at the event horizon (provided that they are vanishing at large distances) and therefore are physically unacceptable. 

A key step in the analysis is the reduction of the perturbation equations to a simple, tractable form. For this purpose, in the case of a four-dimensional black hole, the symmetry of the background space-time is utilized to classify the black hole perturbations into two types, axial (odd) modes and polar (even) modes, according to their behavior under the parity transformation on the two-sphere. To linear order, the perturbation equations for these two types of perturbations are independent, and therefore they reduce to a set of two Schrodinger-type second-order ordinary differential equations, the Reggae-Wheeler equation for axial perturbations and the Zerilli equation \cite{ZerilliLett} \cite{ZerilliD} for polar perturbations. 

In the higher dimensional case, the role of the two-sphere is played by an ($N-$1)-variety: According to their tensorial behavior on the ($N-$1)-variety, perturbations are classified into three types, those of {\it tensor, vector}, and {\it scalar} modes, according to their tensorial behavior on the section of the background space-time. The first of these is a novel type of perturbation that exists only in the higher dimensional case, while the vector- and scalar-type modes correspond, respectively, to the axial and polar modes in the four-dimensional case.

In order to investigate the asymptote of a perturbation of higher dimensional space-times, one must study the asymptote of all the types of perturbations mentioned above. However, besides from Ishibashi and Kodama's work, all analysis made to this time has been carried out for only tensor perturbations \cite{GibbonsHartnoll}, and the situation for vector and scalar perturbations has not yet been investigated: The asymptote of a perturbation of space-times in higher dimensions is not yet established. The treatment of scalar perturbations\footnote{The term ``scalar perturbation" is often used in reference to a perturbation of a free scalar field in works involving quasi-normal mode analysis of black holes. This is conceptually quite different from the scalar perturbations considered in this paper, which are modes of space-time perturbations of a vacuum space-time. Furthermore, the effective potentials for these two cases are different. Hence, although the quasi-normal mode analysis of a test scalar field \cite{Cardoso} \cite{HorowitzHubeny} \cite{Konoplya} \cite{Molina} may shed some light on the problem of the asymptote of a perturbation of static homogeneous space-times in higher dimensions, an exact analysis of the problem based on the formulation given in this paper is necessary. } is the most difficult part of this analysis, because the linearized Einstein equations for a scalar perturbation are much more complicated than those for a tensor perturbation and do not reduce to the form of a single second-order Schrodinger-type in the higher dimensional case.

Recently, this reduction was carried out by Ishibashi and Kodama \cite{KodamaA} in the background of maximally symmetric black holes with a non-vanishing cosmological constant in an arbitrary number of space-time dimensions. From the result of that work, along with the previously obtained ordinary differential equations for tensor and vector perturbations \cite{KodamaIshibashiSeto}, we have the full set of equations for the perturbations of higher dimensional maximally symmetric black holes in the form of a single self-ad-joint second-order ordinary differential equation --- which we refer to as the {\it master equation} --- for each tensorial type of perturbation. These master equations of course, can be applied to the perturbation analysis of higher dimensional space-times. Let us mention, furthermore, that these master equations have been derived in the framework of the gauge-invariant formalism for the perturbations, and for this reason, they do not involve the problem of the choice of gauge.

In the gauge invariant approach we can isolate the {\it configuration variables}, which carry the descriptive arbitrariness of the theory, from the {\it observables}, which are gauge invariant quantities providing a coordination of the reduced phase space of general relativity, and are subjected to hyperbolic (and therefore {\it causal} in the customary sense) evolution equations. In physics the {\it Hole Argument} is an aspect of the fact that also Einstein's theory is interpreted as a gauge theory.

Recall that we consider as background the Cartesian product of a two-dimensional orbit space ${\mathcal N}\sp2$ and a horizon with $N-$1 compact dimensions, where 1+$N$ is the total space-time dimension and $N\ge$ 4 (in order to avoid space-times with one or less compact spatial dimensions where the presence of a massive source is inconsistent with asymptotic flatness).

In this background one expects two linearly independent master solutions. When the size of the radial variable is small (the order of the size of the horizon) one expects that the solutions closely resemble a (1+$N$)-dimensional static homogeneous space-time, while as one increases the distance one expects that at some point the perturbed solution will be flat.

As mentioned above, we have already obtained the master equation for each tensorial type of perturbation in the background space-time. Our main focus in this section is therefore to prove that for each harmonic mode of the horizon there are two solutions that behave similarly at large. In the basic mode, this behavior is ${{(-{r^2}+{t^2})}^{\frac{1-N}{2}}}$. In particular, an eigentensor of the Lichnerowicz operator in a (possibly higher dimensional) black hole with Zero eigenvalue decays at infinity in this way. These solutions occur in an integral form. In particular, there exist no unstable solutions to the master equation with physically acceptable initial data.

Ishibashi and Kodama showed that there exist no well-behaved perturbations that are regular everywhere outside the event horizon. This is consistent with the uniqueness \cite{GibbonsLett} of a certain class of black holes in an asymptotically flat background. Although our main interest is in the standard higher dimensional black holes, that is, asymptotically flat,\footnote{It has been pointed out, however, that the concept of asymptotic flatness itself involves some subtle problems in higher dimensions, especially in the case of an odd number black hole dimensions \cite{Hollands}.} spherically symmetric static vacuum solutions in higher dimensions, we also investigate the asymptote of a perturbation of other maximally symmetric static homogeneous space-times with a non-vanishing cosmological constant with respect to tensor- and vector-type perturbations.

There is a reason to expect a good analytic control of static homogeneous space-time perturbations even if we do not have a complete analytic solution since we have a good approximation: the space-time is expected to resemble closely a (1+$N$)-dimensional static homogeneous black hole \cite{Tangherlini}.

The motivations for this research are first to obtain a theoretical description of a small perturbation which is important on its own right, and second to gain understanding of perturbation physics through combination with analytical work. The symbiosis with analytical work comes close to serve as a partial substitute of experiments (which are sorely absent in this field): the analytics are essential for understanding static solutions close to perturbation where the approximation is expected to hold, and serves to formulate the aims and methods of the theory. As it turns out, the black holes obtained numerically show only a single multi-pole mode perturbation to their horizon \cite{SorkinKolPiran} \cite{KudohWiseman}, and that lends some hope that the analytic approximation would retain some validity for static homogeneous space-times as well.

Our objective is to describe the perturbation process (to first order). Writing down the equations of motion and separating the horizontal variables we find that a Ric flatness condition in the horizontal directions yields a relation among the radial functions which is similar to a trace condition and allows us to simplify the fields. After substitution one can express the fields in terms of first and second derivatives. We are left with a second order tensor differential equation which reduces to a single second order ordinary differential equation in the radial direction, for each metric function (and for each harmonic mode) from which the whole metric may be recovered. This is the master equation, which after a change of variables simplifies further. It would be nice to have a deeper understanding why these reductions were to be expected.

The master equation belongs to the class of {\it Fuchsias equations}, where Fuchsias means that the equation has only regular singular points on the complex sphere which includes infinity, and means that there are exactly three such points. As the {\it hyper-geometric} case of 3 regular singularities there is a general solution to the equation, and several methods are available. Our case is a rather special case of the hyper-geometric equation. In this case, it turns out that the solutions can be written in terms of an {\it integral transformation}, and it would be nice to understand why that had to be the case. Interestingly, we observe that in some of the relevant cases these integral transformations simplify further to {\it explicit functions}, and in particular, all relevant solutions are explicit functions at infinity (solutions which are of even multi-pole number and are regular at the horizon).

We started working out this problem before we were aware of the results of Kodama and Ishibashi \cite{KodamaA} \cite{Ishibashi} \cite{KodamaStability}. Even though their papers were published more than three years ago, we continued independently after learning about them the formalism of gauge invariant perturbation theory, and the various changes of variables which are employed there. We were able to do so and actually found a somewhat similar master equation. Yet the final reduction of our master equation to a hyper-geometric one was motivated by those papers.

In Subsection \ref{Subseccion 4.1} we briefly overview the present status of the investigations of this problem and discuss what kind of new information can be obtained by the linear perturbation theory. In Subsection \ref{Subseccion 4.2} we discuss properties of the eigentensors of the Lichnerowicz Laplacian for a space-time with applications to the asymptote of a linear perturbation of higher dimensional black holes. In Subsection \ref{Subseccion 4.3} we determine the linearized perturbations of the space-time. We have actually solved the perturbation equation with initial data of a static symmetrical homogeneous space-time in Section \ref{Seccion 2}. We apply this to perform the analysis of the asymptote of a perturbation for tensor, vector and scalar perturbations separately.
\subsection{Higher-dimensional static homogeneous\\
space-times}
\label{Subseccion 4.1}
In this subsection we write the static Einstein equations in a form that will be convenient for expansion around a background space-time. This type of expansion is known as {\it gauge invariant perturbation} and we follow here the usual conventions for this type of expansion.

We denote in this subsection the (1+$N$)-dimensional space-time metric by $g_{i j}$. For the gauge invariant perturbation it is convenient to write the Ric tensor in the following form \cite{FockThe}
\begin{equation}
{\hbox{Ric}_{\hbox{\textgoth g}}}_{i j}=-\frac12 g _{i k} g _{j l} g \sp{m n} \frac{{\partial\sp2} g\sp{l k}}{\partial x\sp m\partial x\sp n}+\Gamma \sp{m n}_i\Gamma_{j,m n} -\Gamma_{i j},
\end{equation}
where $\Gamma_{i, m n}$ and $\Gamma_i\sp{m n}$ are the Christoffer symbols of the first and the second kind, respectively, and in addition one defines
\begin{eqnarray}
\Gamma_{i j}&\equiv&\frac12\left(g_{i l} \frac{\partial\Gamma\sp l}{\partial x\sp j}+g_{j k} \frac{\partial\Gamma\sp k}{\partial x\sp i}-g_{i l} g_{j k} \frac{\partial g\sp{l k}}{\partial x\sp m}\Gamma\sp m\right), \\
\Gamma\sp j&\equiv&g \sp{m n}\Gamma\sp j_{m n}.
\end{eqnarray}

The first step in this procedure is to look at the linearized equations valid for weakly gravitating space-times. The metric is taken to be
\begin{equation}
g_{i j}+b_{i j},
\end{equation}
where we denote the perturbation metric by $b_{i j}$. We have (1+$N$)(2+$N$)/2 metric functions $b_{i j}$ which are functions of ($r$, $t$, $\chi$). The linearized field equations become
\begin{equation}
\Delta_{\hbox{\textgoth g}}b_{i j}+{\hbox{\rm Ric}_{\hbox{\textgoth g}}}_{i k}b_j\sp k+{\hbox{\rm Ric}_{\hbox{\textgoth g}}}_{j k}b_i\sp k-2{\hbox{\rm Riemann}_{\hbox{\textgoth g}}}_{i k j l}b\sp{k l}=G_{i j}=8\pi G_{1+N} T_{i j},
\end{equation}
where $\Delta_{\hbox{\textgoth g}}$ is the Laplace operator in the metric {\textgoth g}.

We open this subsection with one comment. At higher orders in the perturbation procedure the form of the equations is dominated by the linearized equations and is given by
\begin{equation}
\Delta_{\hbox{\textgoth g}}g\sp{(m),i j}=F(\hbox{\textgoth g}, \partial \hbox{\textgoth g}),
\end{equation}
where ($m$) is the order under study and $F(\hbox{\textgoth g}, \partial \hbox{\textgoth g})$ are source terms which are quadratic, at least, in lower order metric components and their derivatives.
\subsubsection{Generalized static solution}
It is quite easy to find the higher dimensional counterparts of 4D static solutions, such as the Tangherlini solution \cite{Tangherlini}. In reality, it is also possible to find a slightly more general family of solutions for the Einstein-Maxwell system with cosmological constant by requiring that the space-time metric and the electromagnetic tensor ${\mathcal F}$ can be expressed in the form (\ref{2}) and
\begin{equation}
\label{15}
{\mathcal F}=\frac{1}{2}E_0(\epsilon_{00}d t\wedge d t+\epsilon_{01}d t\wedge d r+\epsilon_{11}d r\wedge d r).
\end{equation}
The Einstein equations and Maxwell equations determine the two dimensional metric and the electric field $E_0$ as \cite{KodamaProg} \cite{Birmingham}
\begin{equation}
\label{16}
g_{11}=\frac{1}{g_{00}}, \quad E_0=\frac{Q}{r\sp{N-1}}, 
\end{equation}
where
\begin{equation}
g_{00}(r)=\frac S{(N-2)(N-1)}-\frac R{N(1+N)} r\sp2-\frac{\rho\sp{N-2}}{r\sp{N-2}}+\frac{\kappa\sp2Q\sp2}{(N-2)(N-1)r\sp{2(N-2)}}.
\end{equation}
This formula includes the solution with cosmological constant, which is mentioned above.

In the special case in which $M\sp{N-1}$ is the sphere $S\sp{N-1}$, the space-time becomes static and spherically symmetric. In particular, for
\begin{equation}
\frac{S}{(N-2) (N-1)}=1, R=0\quad\hbox{and}\quad Q=0,
\end{equation}
this solution coincides with the Tangherlini solution, and for
\begin{equation}
\frac{S}{(N-2) (N-1)}=1\quad\hbox{and}\quad R=0,
\end{equation}
it gives a higher dimensional counter-part of the Reissner-Nostrum solution. More generally, when $M\sp{N-1}$ is a constant curvature space, the solution is invariant under SO($N$), ISO($N-$1) and SO($N-$1, 1) for
\begin{equation}
\frac{S}{(N-2) (N-1)}=1, 0, -1,
\end{equation}
respectively.

In general, this solution has a naked singularity or does not have a horizon. As shown in Ref. \cite{KodamaProg}, it gives a regular black hole only when the parameters $Q$, $R$ and $S$ are in special regions. In this case, the Einstein space $M\sp{N-1}$ describes a spatial section of the static homogeneous space-time horizon and at the same time the spatial infinity.
\subsubsection{Uniqueness about static solutions}
In the static case with $R$=0, the uniqueness about asymptotically flat regular static solutions are now established in higher dimensions as well: all regular solutions are exhausted by the Tangherlini solution for the vacuum system \cite{Hwang}, Reissner-Nostrum solution given by (\ref{2}), (\ref{15}) and (\ref{16}) with $\frac{S}{(N-2) (N-1)}$=1 and $R$=0 in the non-degenerate case \cite{GibbonsLett} and the higher dimensional Majumdar-Papapetrou solutions in the degenerate case for the Einstein-Maxwell system \cite{Rogatko}. It is also proved that the Gibbons-Mada solution and the Tangherlini solution are the only asymptotically flat regular static symmetrical homogeneous solutions for the Einstein-Maxwell-Dilator system and for the Einstein-Harmonic-Scalar system, respectively \cite{GibbonsD} \cite{RogatkoClass}. However, for $R\neq$ 0, nothing is known about the uniqueness, although it was pointed out in Ref. \cite{AndersonChrusciel} that the approach developed by Anderson may also apply to higher dimensional cases with $R<$ 0.

In the rotating case, the higher dimensional counterpart of the Kerr solution is also known. It is given by the Myers-Perry solution \cite{Myers}, which has a horizon and is asymptotically flat. In contrast to the static case, however, this is not the unique asymptotically flat rotating regular solution in higher dimensions, although the uniqueness holds for super-symmetric static homogeneous black holes in the 5-dimensional minimal super-gravity model \cite{Reall}. This is because the horizon topology need not be given by the sphere in higher dimensions as discussed in Ref. \cite{Cai}. In fact, Emparan and Reall \cite{Emparan} found an asymptotically flat and rotating regular static symmetrical homogeneous solution in five dimensions whose black hole surface is homeomorphic to $S\sp2\times S\sp1$. Thus, at present, we have two different families of asymptotically flat rotating regular solutions with different horizon topologies. Since a more complicated horizon topology is allowed in dimensions greater than 5, it is highly probable that other families of regular solutions exist. However, near the static limit, it is likely that a uniqueness theorem holds.
\subsection{Fluctuations about static homogeneous space-times}
\label{Subseccion 4.2}
Fluctuations about static homogeneous space-times relate to their stability. In particular, if the cosmic censorship hypothesis  does not hold, a regular stationary static homogeneous space-time solution will be unstable against a generic perturbation, because the formation of a naked singularity must be generic \cite{Waldgr}.

In reality, many of the standard static homogeneous space-time solutions are known to be perturbatively unstable \cite{Gross}.

In contrast to static homogeneous space-times, the fluctuations about rotating black holes are established only for the Kerr solutions \cite{Whiting}. We do not discuss the fluctuations about these rotating static solutions in this article.

In the proofs of the fluctuations about these static solutions, a key role was played by the fact that the basic equations for the perturbations can be reduced to a single second-order ordinary differential equation of the Schrodinger type, which is often called a master equation. Now, we explain how such a master equation is derived and how the asymptote of a perturbation is calculated in terms of it.
\subsubsection{Tensorial decomposition of perturbations}
In space-times of dimension 1+$N$, the linearizing of the Einstein equations yields the following equations for the space-time perturbation $b_{i j}$=$\delta g_{i j}$:
\[
(\hbox{Lichnerowicz}_{\hbox{\textgoth g}}b)_{i j}-\nabla_i\nabla_j b+2\nabla_{(i}\nabla\sp k b_{j)k}
\]
\begin{equation}
+(-\nabla\sp k\nabla\sp l b_{k l}+\Delta_{\hbox{\textgoth g}} b+\hbox{Ric}_{\hbox{\textgoth g}}\sp{k l}b_{k l})g_{i j}-\frac{2R}{1+N}b_{i j}=2\kappa\sp2\delta T_{i j}, 
\end{equation}
where Lichnerowicz$_{\hbox{\textgoth g}}$ is the Lichnerowicz operator defined by
\begin{equation}
(\hbox{Lichnerowicz}_{\hbox{\textgoth g}}b)_{i j}=\Delta_{\hbox{\textgoth g}}b_{i j}+{\hbox{\rm Ric}_{\hbox{\textgoth g}}}_{i k}b_j\sp k+{\hbox{\rm Ric}_{\hbox{\textgoth g}}}_{j k}b_i\sp k-2{\hbox{\rm Riemann}_{\hbox{\textgoth g}}}_{i k j l}b\sp{k l}.
\end{equation}

In order to analyze the behavior of the perturbations on the basis of these equations, the following two technical problems have to be resolved. Firstly, the perturbation variables $b_{i j}$ contain unphysical gauge degrees of freedom that should be eliminated. Second, these perturbation equations are a coupled system of a large number of equations and are quite hard to solve directly.

In the case of our interest, these problems can be resolved with the help of the tensorial decomposition and the gauge-invariant formulation \cite{KodamaIshibashiSeto} \cite{Bardeen} \cite{KodamaSasaki}. First, the metric of a static homogeneous space-time solution to the vacuum or electron-vacuum system in 1+$N$ dimensions can be written in terms of the two-dimensional orbit space ${\mathcal N}\sp2$ as (\ref{2}). On this background, the basic perturbation variables $b_{i j}$ can be grouped into three sets, 
\begin{itemize}
\item0$\le i$, $j\le$1, 
\item0$\le i\le$1 and 2$\le j\le N$+1, and
\item2$\le i$, $j\le N$+1, 
\end{itemize}
according to their tensorial transformation behavior on $M\sp{N-1}$. Among these, $b_{i j}$ for 0$\le i\le$1 and 2$\le j\le N$+1, can be further decomposed into the gradient part $b_i$ and the divergence-free part $b_{i j}\sp{(1)}$ as
\begin{equation}
b_{i j}=\hat D_j b_i+b_{i j}\sp{(1)};\quad\hat D_j b_{i j}\sp{(1)}=0, 
\end{equation}
where $\hat D_j$ is the co-variant derivative with respect to the metric $h_{i j}$ on $M\sp{N-1}$. Similarly, $b_{i j}$ for 2$\le i$, $j\le N$+1, can be decomposed into the trace part and trace-free part as
\begin{equation}
b_{i j}=b_L h_{i j}+b_{T i j};\quad b_{T j}\sp j= 0, 
\end{equation}
and the trace-free part $b_{T i j}$ can be further decomposed into the scalar part $b_T\sp{(0)}$, the divergence-free vector part $b_{T j}\sp{(1)}$ and the divergence-free and trace-free part $b_{T i j}\sp{(2)}$ as
\begin{equation}
b_{T i j}=\left(\hat D_i\hat D_j+\frac{1}{N-1}h_{i j} \Delta_{\hbox{\textgoth h}}\right)b_T\sp{(0)}+2 \hat D_{(i}b_{T j}\sp{(1)})+b_{T i j}\sp{(2)};
\end{equation}
\begin{equation}
\hat D\sp j b_{T j}\sp{(1)}=0, \quad\hat D\sp j b_{T i j}\sp{(2)}=0.
\end{equation}

By this decomposition, the linearized Einstein equations are decomposed into three decoupled sets of equations, each of which contains only variables belonging to one of the three sets of variables, the scalar-type variables ($b_{00}$, $b_{01}$, $b_{11}$, $b_i$, $b_L$, $b_T\sp{(0)}$), vector-type variables ($b_{i j}\sp{(1)}$, $b_{T j}\sp{(1)}$), and tensor-type variables ($b_{T i j}\sp{(2)}$). Further, since the co-variant derivative $\hat D_j$ always appears in the combination $\Delta_{\hbox{\textgoth h}}$=-$\hat D_i\hat D\sp i$ in the perturbation equations, the expansion coefficients of these variables in terms of harmonic tensors on $M\sp{N-1}$, which are denoted as ($\xi_{00}$, $\xi_{01}$, $\xi_{11}$, $f_i\sp{(0)}$, $H_L$, $H_T\sp{(0)}$), ($f_i\sp{(1)}$, $H_T\sp{(1)}$) and $\xi$ for the scalar-type, vector-type and tensor-type variables, respectively, are mutually coupled among only those corresponding to the same harmonic tensor (see Ref. \cite{GibbonsHartnoll} \cite{KodamaProg} for details of this harmonic expansion). Thus, the perturbation equations can be reduced to sets of equations on the two-dimensional orbit space ${\mathcal N}\sp2$ with small number of entries independent of the space-time dimension. In addition, after this harmonic expansion, we can easily construct a basis for gauge-invariant variables by taking appropriate linear combinations of these two-dimensional variables and their derivatives with respect to our coordinates of ${\mathcal N}\sp2$, and the linearized Einstein equations can be written as differential equations for these gauge-invariant variables \cite{KodamaIshibashiSeto}.
\subsection{Eigentensors of the Lichnerowicz operator}
\label{Subseccion 4.3}
In this subsection we find the linear perturbations for the (1+$N$)-dimensional static symmetrical homogeneous solution and show that an eigentensor of the Lichnerowicz operator is regular at infinity. Reggae and Wheeler \cite{Regge} derived the linear equations that describe small perturbations of the four dimensional black hole, and here we generalize their method for the static case. These perturbations can be interpreted as deviations of the static homogeneous black hole from symmetry due to remote masses. In the case of a sphere, we are interested in the influence of the static homogeneous black hole (images) on itself.

As explained in the introduction the zeroth order correction in the horizon metric is the (1+$N$)-dimensional static symmetrical homogeneous black hole \cite{Tangherlini}. More generally, as our background, we consider the (1+$N$)-dimensional metric {\textgoth g} of the form (\ref{2}) on the Cartesian product of the two-dimensional orbit space ${\mathcal N}\sp2$ and the ($N-$1)-dimensional maximally symmetric space $M\sp{N-1}$, with curvature scalar satisfying $\frac{{S}}{(N-2) (N-1)}$=0, $\pm$1. The constant scalar of curvature $R$ of the space-time is related to the (1+$N$)-dimensional cosmological constant $\Lambda$ as $\Lambda$=2$R$, and the parameter ${{\rho }^{N-2}}/2$ defines the static solution mass. For example, in the case $\frac{{S}}{(N-2) (N-1)}$=1 and $R$=0, this mass is given by
\begin{equation}
\frac{(N-1){{\rho }^{N-2}}{\mathcal A}_{N-1}}{16\pi G_{1+N}}, 
\end{equation}
where ${\mathcal A}_{N-1}$=2$\pi\sp{N/2}$/$\Gamma$[{$N$/2}] is the area of a unit ($N-$1)-sphere, and $G_{1+N}$ denotes the (1+$N$)-dimensional Newton constant. We assumed that ${{\rho }}\ge$ 0 to avoid the appearance of a naked singularity in the background. We are interested in the static region --- called the {\it black hole wedge} --- in which
\begin{equation}
{{-{r^{2-N}}{{\rho }^{N-2}}-\frac{{r^2}{R}}{N (1+N)}+\frac{{S}}{(N-2) (N-1)}}} > 0,
\end{equation}
so that $t$ is a space-like coordinate. Thus, when $R\ge$ 0, the curvature scalar $S$ must be positive, while when $R<$ 0, $\frac{{S}}{(N-2) (N-1)}$ can be 0 or $\pm$1. The higher dimensional black hole corresponds to the case $\frac{{S}}{(N-2) (N-1)}$=1, ${{\rho }}>$ 0 and $R$=0. 

To study perturbations in this background, it is convenient to introduce harmonic tensors in the ($N-$1)-dimensional symmetric space ($M\sp{N-1}$, $\hbox{\textgoth h}$) and expand perturbations in terms of them, as we see below. The type of the tensor defines the type of perturbation. Combining the expansion coefficients, one can construct the gauge-invariant variables in the two-dimensional orbit space metric spanned by our coordinates $y\sp i$=($t$, $r$). The linearized Einstein equations then reduce to a set of equations for the gauge-invariant quantities. It is found from the Einstein equations that there is a master scalar variable for each tensorial type of perturbation. Consequently, the linearized Einstein equations reduce to a set of three master equations for the master variable. Thorough studies of the gauge-invariant formalism for perturbations and the derivation of the master equation are given in Refs. \cite{KodamaA} and \cite{KodamaIshibashiSeto}.

To determine the eigenvalues of the Lichnerowicz operator we study the solutions of
\begin{equation}
\label{17}
\Delta_{\hbox{\textgoth g}}b_{i j}+{\hbox{\rm Ric}_{\hbox{\textgoth g}}}_{i k}b_j\sp k+{\hbox{\rm Ric}_{\hbox{\textgoth g}}}_{j k}b_i\sp k-2{\hbox{\rm Riemann}_{\hbox{\textgoth g}}}_{i k j l}b\sp{k l}=\lambda b_{i j}, 
\end{equation}
where $b$ is transverse, normalizable and
\begin{equation}
-\frac{b_{00}}{g_{00}}+\frac{b_{11}}{g_{11}}=0,
\end{equation}
\begin{equation}
\sum_{i, j=2}\sp{1+N}h\sp{i j}b_{i j}=0.
\end{equation}
A variant of this problem has been treated by Reggae and Wheeler \cite{Regge}. They investigated the Lorentz version of this problem with $\lambda$=0. Their methods were refined by subsequent workers, Vishveshwara \cite{Vishveshwara}, Zerilli \cite{ZerilliLett}, Press and Teukolsky \cite{185} \cite{193}, Stewart \cite{Stewart}, and Chandrasekhar \cite{Chandrasekhar}. $\lambda$=0 corresponds to a small perturbation of a black hole that remains a classical solution. These authors searched for runaway solutions of the form exp(-$i\omega t$)$\times$(function of spatial variables with $\omega$ complex). They demonstrated that Imaginary$\omega$=0 for all solutions of (\ref{17}) and concluded that black holes were classically stable objects.

We, on the other hand, are interested in solutions to (\ref{17}), where $g_{i j}$ is a static symmetrical homogeneous solution and $\lambda$ is not necessarily Zero. Positive (negative) values of $\lambda$ correspond to stable (unstable) Gaussian fluctuations about our space-time.

By applying the above arguments, we obtain three decoupled sets of equations for the scalar-type, vector-type and tensor-type perturbations. The first two are often called the even modes (or the polar perturbation) and the odd modes (or the axial perturbation), respectively. This equation can be separated in $t$, $r$ and $M\sp{N-1}$ coordinates, and is exhibited in equations (\ref{19}), (\ref{20}) and (\ref{21}).

We then follow the approach of Reggae and Wheeler \cite{Regge} and divide the space of eigenfunctions. Since the background is static, by the Fourier transformation with respect to the time coordinate $t$,
\begin{equation}
b_{i j}(t, r,...)=\left\{\begin{array}{ll}
\tau_{i j}(t)\xi_{i j}(r)E_{i j}, &\hbox{if }0\le i\le1, 0\le j\le 1+N, i\le j\\
\tau(t)\xi(r)\beta_{i j}, &\hbox{if }2\le i, j\le 1+N
\end{array}\right..
\end{equation}
where
\begin{itemize}
\item
\begin{equation}
-\frac{d\sp2\tau}{d t\sp2}=\omega\sp2\tau, 
\end{equation}
\item
\begin{equation}
-\frac{d\sp2\tau_{i j}}{d t\sp2}=\omega_{i j}\sp2\tau_{i j}, \quad0\le i\le1, \quad0\le j\le 1+N, \quad i\le j, 
\end{equation}
\item
\begin{equation}
\Delta_{\hbox{\textgoth h}}E_{i j}=e_{i j}E_{i j}, \quad0\le i\le1, \quad0\le j\le 1+N, \quad i\le j, 
\end{equation}
and
\item
\begin{equation}
\label{18}
\Delta_{\hbox{\textgoth h}}\beta_{i j}+{\hbox{\rm Ric}_{\hbox{\textgoth h}}}_{i k}\beta_j\sp k+{\hbox{\rm Ric}_{\hbox{\textgoth h}}}_{j k}\beta_i\sp k-2{\hbox{\rm Riemann}_{\hbox{\textgoth h}}}_{i k j l}\beta\sp{k l}=\mu\beta_{i j},
\end{equation}
for 2$\le i$, $j\le N$+1.
\end{itemize}
It turns out that the eigenvalues equation of the Lichnerowicz operator simplify. The linearized vacuum Einstein equations can be brought to a simplified form \cite{WaldGeneral}. Because our background space-time is static, with a Killing parameter $t$ in the black hole wedge, substitution of these forms into (\ref{17}), together with the conditions of trace-free and transversality applied to $b_{i j}$, leads to sets of decoupled ordinary differential equations for the scalar-type, vector-type and tensor-type perturbations in the background space-time. The corresponding set of equations can be easily casted into a single second-order ordinary differential master equation of the Schrodinger type, as first shown by Reggae and Wheeler \cite{Regge}. It was shown by Zerilli that such reduction exists for the scalar-type perturbation of a black hole \cite{ZerilliLett}. Although these authors derived these master equations by the gauge-fixing method, the master variable can be written in terms of gauge-invariant variables\footnote{The problem can be characterized by a single dimensionless variable, $x$=$\rho r\sp{-1}$.} \cite{KodamaA}. Thus, the linear fluctuations about a black hole are determined by the behavior of the functions $\xi$'s for each mode.

In the subsequent subsections, we examine tensor, vector and scalar perturbations separately, briefly reviewing the space-time perturbations and the master equation.
\begin{lemma}
The parts of the Riemann tensor are
\[
{\hbox{\rm Riemann}_{\hbox{\textgoth g}}}_{i j k l}=
\]
\[
g_{00}\left[\frac{R g_{11}}{1+N}-\frac{N-1}r\left(\frac{g_{11}'}{2g_{11}}+\frac1r\right)\right]\times
\]
\[
\times(\delta_i\sp0\delta_j\sp1\delta_k\sp0\delta_l\sp1-\delta_i\sp0\delta_j\sp1\delta_k\sp1\delta_l\sp0-\delta_i\sp1\delta_j\sp0\delta_k\sp0\delta_l\sp1+\delta_i\sp1\delta_j\sp0\delta_k\sp1\delta_l\sp0)
\]
\[
+g_{00}\left(\left[\frac{R}{1+N}r\sp2-\frac{1}{g_{11}}\left(\frac{r g_{11}'}{2g_{11}}-(N-3)\right)\right]-\frac{S}{N-1}\right)
\]
\[
\times(h_{j l}\delta_i\sp0\delta_k\sp0-h_{j k}\delta_i\sp0\delta_l\sp0+h_{i k}\delta_j\sp0\delta_l\sp0-h_{i l}\delta_j\sp0\delta_k\sp0)
\]
\[
-\left(\frac{r g_{11}'}{2g_{11}}+1\right)(-h_{j l}\delta_i\sp1\delta_{k}\sp1+h_{j k}\delta_i\sp1\delta_{l}\sp1-h_{i k}\delta_{j}\sp1\delta_{l}\sp1+h_{i l}\delta_{j}\sp1\delta_{k}\sp1)
\]
\begin{equation}
+\frac{{r}\sp2}{g_{11}}(h_{i l}h_{j k}-h_{i k}h_{j l})+r\sp2{\hbox{\rm Riemann}_{\hbox{\textgoth h}}}_{i j k l}.
\end{equation}
\end{lemma}
\begin{lemma}
We have
\[
{\hbox{\rm Ric}_{\hbox{\textgoth g}}}_{i k}b_j\sp k+{\hbox{\rm Ric}_{\hbox{\textgoth g}}}_{j k}b_i\sp k-2{\hbox{\rm Riemann}_{\hbox{\textgoth g}}}_{i k j l}b\sp{k l}=
\]
\begin{equation}
\left\{\begin{array}{l}
\left(\frac{2R}{N}+(N-2)(N-1){{\rho }^{-2+N}}r^{-N}\right)b_{i j}\\
\left(\frac{2R}{N}-(N-2){{\rho }^{-2+N}}r^{-N}\right)b_{i j}\\
\left(\frac{2R}{N}-\frac{2S}{(N-2)(N-1)r\sp2}+2{{\rho }^{-2+N}}r^{-N}\right)b_{i j}-\frac{2}{r\sp2}{\hbox{\rm Riemann}_{\hbox{\textgoth h}}}_{i k j l}b\sp{k l}
\end{array}
\right.,
\end{equation}
if
\begin{equation}
\left\{\begin{array}{l}
0\le i, j\le1\\
0\le i\le1\hbox{ and }2\le j\le 1+N\\
2\le i, j\le 1+N
\end{array}
\right..
\end{equation}
\end{lemma}
Set
\begin{equation}
L_\lambda=-\frac{1}{r\sp{N-1}\sqrt{g_{00}g_{11}}}\frac{d}{d r}r\sp{N-1}\sqrt{\frac{g_{00}}{g_{11}}}\frac{d}{d r}+\frac{2R}{N}-\lambda.
\end{equation}
\subsubsection{Tensor perturbations}
We begin by considering tensor perturbations, which are given by
\begin{equation}
b_{i j}=\left\{\begin{array}{ll}
0,&0\le i\le1, 0\le j\le 1+N, i\le j\\
\xi\beta_{i j},&2\le i, j\le 1+N
\end{array}
\right.,
\end{equation}
where the tensor $\beta_{i j}$ is defined as solution to the eigenvalue problem on the ($N-$1)-variety, i.e., Ref. (\ref{18}) with
\begin{equation}
\beta\sp i_i=0,\quad \hat D_j \beta\sp j_i=0. 
\end{equation}
The eigenvalues $\mu$ are given by $l$($l$+$N-$2)$-$2+2($N-$1) for $\frac{{S}}{(N-2) (N-1)}=1$, (the hyper sphere) and form a continuous set of non-negative values for $S\le$ 0. Note that, in the four-dimensional black hole ($N$=3), there are no corresponding modes, as there are no tensors of this type on a 2-sphere.

It can immediately be seen that the expansion coefficient $f$ is itself gauge independent, and it can be taken as the master variable for a tensor perturbation. The master equation follows from the vacuum Einstein equations
\begin{equation}
\label{19}
\left[L_\lambda+\left(\mu-\frac{2S}{N-2}\right)\frac{1}{r\sp2}+2{{\rho }^{-2+N}}r^{-N}+\frac{\omega\sp2}{g_{00}}\right]\xi=0.
\end{equation}
\subsubsection{Vector perturbations}
Next, consider vector perturbations, which are given by
\begin{equation}
b_{i j}=\left\{\begin{array}{ll}
0,&0\le i, j\le1\hbox{ or }2\le i, j\le 1+N\\
\xi_{i j} E_{i j},&0\le i\le1\hbox{ and }2\le j\le 1+N
\end{array}
\right.,
\end{equation}
where the vector harmonics are introduced as the solutions of
\begin{equation}
(\Delta_{\hbox{\textgoth h}}-e_{i j})E_{i j}=0,
\end{equation}
with eigenvalues $e_{i j}$ that are given by $l_{i j}(l_{i j}+N-2)$ with $l_{i j}\ge$ 0 for
\begin{equation}
\frac{{S}}{(N-2) (N-1)}=1,
\end{equation}
and form a continuous set of non-negative values for $S\le$ 0.

A gauge-invariant variable $\xi_{i j}$ can be constructed. Substituting this into the Einstein equations, we can obtain the master variable $\xi_{i j}$.

The radial part of the Einstein equations can then be reduced to
\begin{equation}
\label{20}
\left(L_\lambda+\frac{e_{i j}}{r\sp2}-(N-2){{\rho }^{-2+N}}r^{-N}+\frac{\omega_{i j}\sp2}{g_{00}}\right)\xi_{i j}=0.
\end{equation}
\subsubsection{Scalar perturbations}
We now examine scalar-type perturbations in the background space-time. These perturbations are given by
\begin{equation}
b_{i j}=\left\{\begin{array}{ll}
\xi_{i j} E_{i j},&0\le i, j\le1\\
0,&2\le j
\end{array}
\right.,
\end{equation}
where the scalar harmonics $E_{i j}$ are defined by
\begin{equation}
(\Delta_{\hbox{\textgoth h}}-e_{i j})E_{i j}=0,
\end{equation}
with the eigenvalues $e_{i j}$ given by $l_{i j}(l_{i j}+N-2)$ for $\frac{{S}}{(N-2) (N-1)}$=1 and forming a continuous set of non-negative values for $\frac{{S}}{(N-2) (N-1)}\le$ 0.

We find that the linearized Einstein equations can be reduced to the form:
\begin{equation}
\label{21}
\left(L_\lambda+\frac{e_{i j}}{r\sp2}+(N-2)(N-1){{\rho }^{-2+N}}r^{-N}+\frac{\omega_{i j}\sp2}{g_{00}}\right)\xi_{i j}=0. 
\end{equation}
\subsubsection{The field equations}
We have now 2(1+$N$) radial fields $\xi$($r$), $\xi_{00}$($r$)..., $\xi_{0(1+N)}$($r$), $\xi_{11}$($r$)...,\\$\xi_{1(1+N)}$($r$) defined in (\ref{19}), (\ref{20}), (\ref{21}) for each mode,
\begin{eqnarray}
b_{i j} d x\sp i d x\sp j&=&\xi_{00}E_{00}d t\sp2+\xi_{01}E_{01}(d t d r+d rd t)+\xi_{11}E_{11}d r\sp2\nonumber\\
&&+\sum_{0\le i\le1, 2\le j\le 1+N}\xi_{i j}E_{i j}(d x\sp i d x\sp j+d x\sp j d x\sp i)+\xi\beta.
\end{eqnarray}

Recently, Kodama and Ishibashi have shown that for a static charged homogeneous black hole in higher dimensions represented by (\ref{2}), (\ref{15}) and (\ref{16}), the perturbation equations can be reduced to decoupled second-order master equations of the Schrodinger type, as in four dimensions \cite{KodamaA} \cite{KodamaProg}. These master equations in higher dimensions, however, have some new features. First, there exists no simple relation between vector and scalar perturbations like the scalar-vector correspondence in four dimensions, for $d$=1+$N>$4. This implies that stabilities for scalar and vector perturbations should be studied separately. Second, there exist tensor perturbations for $d$=1+$N>$4.

Now, let us see how these new features affect the fluctuations about static homogeneous space-times in higher dimensions.

When $M\sp{N-1}$ is a generic Einstein space, the $M\sp{N-1}$-coordinates appear in the perturbation equations only through the Lichnerowicz operator
\begin{equation}
(\hbox{Lichnerowicz}_{\hbox{\textgoth h}}\beta)_{i j}=\Delta_{\hbox{\textgoth h}}\beta_{i j}+{\hbox{\rm Ric}_{\hbox{\textgoth h}}}_{i k}\beta_j\sp k+{\hbox{\rm Ric}_{\hbox{\textgoth h}}}_{j k}\beta_i\sp k-2{\hbox{\rm Riemann}_{\hbox{\textgoth h}}}_{i k j l}\beta\sp{k l}
\end{equation}
Hence, by expanding tensor perturbations in terms of the eigentensors $\beta$ of $\hbox{Lichnerowicz}_{\hbox{\textgoth h}}$, 
\begin{equation}
\hbox{Lichnerowicz}_{\hbox{\textgoth h}}\beta=\mu\beta ;\quad \beta_i\sp i=0, \quad\hat D\sp j\beta_{i j}= 0, 
\end{equation}
we obtain a single decoupled equation for $\xi$, which can be easily put into the form (\ref{19}).
\subsubsection{Analysis of the asymptote of a perturbation}
\label{Seccion 3}
We have solved the master equation with initial data of a static symmetrical homogeneous space-time (\ref{13}).

Consider the function
\begin{equation}
f(r)=
\left\{\begin{array}{ll}
(N-2)(N-1){{\rho }^{-2+N}}r^{-N},&0\le i, j\le1\\
-(N-2){{\rho }^{-2+N}}r^{-N},&0\le i\le1\hbox{ and }2\le j\le 1+N\\
-\frac{2S}{(N-2)(N-1)r\sp2}+2{{\rho }^{-2+N}}r^{-N},&2\le i, j\le 1+N
\end{array}
\right.,
\end{equation}
in (\ref{3}). If
\[
\lambda=\frac{2R}{N},
\]
then our universal decay law (-$r^2+t^2)^{(1-N)/2}$ of the field behavior at infinity in space-time (\ref{13}) again agrees with Ref. \cite{Yoshida} because of Equations (\ref{9}) and (\ref{10}).
\section{The Green's function of the Klein-Gordon\\equation in Schwarzschild coordinates}
The present section deals essentially with the field theory in curved space-time by studying a class of ``exact" (integral transforms) solutions to the massive Klein-Gordon field equation in the background of a Schwarzschild-Tangherlini black hole (the higher-dimensional Schwarzschild solution) in an arbitrary number $1+N$ of space-time dimensions.

We again use the relations Equations (\ref{9}) and (\ref{10}) in this study. They are not exact solutions since it is assumed that the amplitude is so small that its contribution to the energy content can be neglected. Some justifications for this study are presented.

We also compute and discuss the behavior of the Green's function of the Klein-Gordon equation for a free scalar field in the background of a curved higher dimensional spherically symmetric Schwarzschild black hole with coordinates coupled to the curvature of the background space-time \cite{Schonberg}. The Green's function is a sum on the harmonic modes of the sphere. We consider the first term in Subsection \ref{Subseccion 5.6}. This term is a double integration on the energy spectrum and the momentum of the particle. Far from the horizon, we can approximate the double integration by an integration on a line defined by the relation of energy and momentum of a free particle. From here, we derive the Yukawa potential (\ref{29}) in our formalism.

Consider the evolution of a massive scalar field $\phi$ in the background described by (\ref{2}). The evolution is governed by the curved space Klein-Gordon equation
\begin{equation}
\label{22}
\left[\frac1{\sqrt{-g}}\partial_\alpha\sqrt{-g}g\sp{\alpha\beta} \partial_\beta-m\sp2\right]\phi= 0,
\end{equation}
where we denote by $g$ the determinant of the metric. The metric appearing in (\ref{22}) should describe the geometry referring to both the black Brana and the scalar field, but if we consider that the amplitude of $\phi$ is so small that its contribution to the energy content can be neglected, then the metric (\ref{2}) should be a good approximation to $g_{\alpha \beta}$ in (\ref{22}). We shall thus work in this perturbation approach. It turns out that it is possible to simplify considerably equation (\ref{22}) if we separate the horizon variables from the radial and time variables, as is done in four dimensions \cite{Brill} \cite{Teukolsky}. For higher dimensions we follow \cite{Ida}.

Consider a number $\xi$. The evolution of a minimally coupled scalar field $\phi$ is described by the massive Klein-Gordon equation (\ref{3}) \cite{Gan} \cite{Cao} \cite{Wazwaz} \cite{Tian} \cite{Chen}, with
\begin{equation}
f=m^2+R\xi,
\end{equation}
where (\ref{3}) is a partial differential operator that contains information about the initial shape of the wave packet at $t$=0. This formula also includes the cosmological constant. The explicit form of the operator (\ref{3}) is the simplest in this general setting.
 
The whole section is based on an integral solution of the Klein-Gordon equation. In Section \ref{Seccion 2}, the linear perturbation equation is derived and solved. Let's find a relation of energy and momentum of a particle with mass in a hyper black hole. We use (\ref{10}) and calculate
\begin{eqnarray}
\omega(k)&=&{\sqrt{{k^2}+{m^2}}}-\frac{(2 {k^2}+{m^2}) {r^{2-N}} {{\rho }^{-2+N}}}{2 {\sqrt{{k^2}+{m^2}}}}+o((\rho/r)\sp{N-1})\nonumber\\
&=&{\sqrt{{k^2}+{m^2}}}-\frac{1}{2} m {r^{2-N}} {{\rho }^{-2+N}}+o(k\sp2)+o((\rho/r)\sp{N-1}).
\end{eqnarray}
The first term is the energy of a free particle. \\
The second term is the gravitational potential energy between the masses $m$ and $\frac{1}{2}{{\rho }^{-2+N}}$.
\subsection{The Green's function}
To solve the inhomogeneous version of the differential equation (\ref{3}), we must construct a Green's function out of the fundamental system of solutions of the homogeneous equation
\begin{eqnarray}
\label{23}
Z_1(r, t, \chi)&=&{r^{1-\frac{N}{2}}}\sum_e E_e(\chi)g(e)\int_0\sp\infty k^{o}{e\sp{-{i} t \omega(k, r, \lambda)}}{B_o}(r k)d k\\
\label{24}
Z_2(r, t, \chi)&=&{r^{1-\frac{N}{2}}}\sum_e E_e(\chi)g(e)\int_0\sp\infty k^{o}{e\sp{{i} t \omega(k, r, \lambda)}}{B_o}(r k)d k,
\end{eqnarray}
where parameters $\pm\omega$ of the differential equation depend on $k, r, \lambda$ as given by equations (\ref{10}) and
\[
f=m^2+R\xi-\lambda.
\]
With these functions, the Green's function is constructed as
\[
G({r_1, t_1, \chi_1}; r_2, t_2, \chi_2)={(r_1r_2)^{1-\frac{N}{2}}}\sum_e E_e(\chi_1)E_e\sp*(\chi_2)\times
\]
\begin{equation}
\label{25}
\int_{-\infty}\sp\infty\int_0\sp\infty k\sp{2o}\frac{e\sp{{i} (t_1 \omega(k, r_1, \lambda)-t_2 \omega(k, r_2, \lambda))}}\lambda{B_o}(r k_1){B_o}(r k_2)d k d\lambda.
\end{equation}
where the plus-minus sign of $\omega$ is taken depending on the arguments.

We see that $F$($e$, $k$, $\omega$) (\ref{11}) already fixes the Green's function (\ref{25}).

The Green's function (\ref{25}) written above is defined on the Schwarzschild-like space-time, but could be analytically continued to the whole $r<\rho$ space.

Once the Green's function has been determined, it is simple to construct the later-time evolution of the wave-packet from the initial data using equation (\ref{25}). As the eigen-value problem (\ref{3}) is not self-ad-joint, the Green's function (\ref{25}) need not be symmetric in its arguments $r_1$, $t_1$, $\chi_1$ and $r_2$, $t_2$, $\chi_2$. Note that the Green's function is calculated from the $\lambda$-modes and so $Z_1$ and $Z_2$ depend on $\lambda$, but we omitted the third argument in $Z_i$($r$, $t$, $\chi$) for brevity.
\subsubsection{Schwarzschild coordinates}
In Schwarzschild coordinates, (\ref{25}) is simplified. We denote by $P_l$ the Legend re polynomial of degree $l$.

We claim that the expression
\[
G(r_1, t_1, \chi_1; r_2, t_2, \chi_2)=\frac1{4\pi\sqrt{r_1r_2}}\sum_{l=0}\sp\infty(1+2l)P_l(\chi_1\cdot\chi_2)\times
\]
\begin{equation}
\int_{-\infty}\sp\infty\int_0\sp\infty k\sp{1+2 l}\frac{e\sp{{i} (t_1 \omega(k, r_1, \lambda)-t_2 \omega(k, r_2, \lambda))}}{\lambda}{B_{\frac{1}{2}+l}}(r k_1){B_{\frac{1}{2}+l}}(r k_2)d k d\lambda.
\end{equation}
is the Green's function for the Klein-Gordon field in the Schwarzschild space-time and utilize it in the subsequent arguments in essential ways. Actually, this expression does not have the global time translation invariance, in spite of the fact that the background space-time is static.
\subsection{Incoming Wave-packet}
To get back from the frequency $\omega$-interval dependence to the physical time evolution, one performs the inverse transformation
\begin{equation}
\label{26}
\phi(r, t, \chi)={r^{1-\frac{N}{2}}}\sum_e E_e(\chi)\int_{-\infty}\sp\infty e\sp{it\omega}\int_0\sp\infty{B_o}(r k)F(e, k, \omega )d k d\omega.
\end{equation}

The method of the stationary phase deals with the approximate evaluation of Fourier-type integrals
\begin{equation}
f(t)=\int_\alpha\sp\beta F(\omega)\exp[it S(\omega)]d\omega
\end{equation}
for large positive parameter $t$.

Inverse transform integrals (\ref{26}) are precisely of the above type, with phase $S$($\omega$)=$\omega$ and the large parameter $t$ being the time coordinate. So the stationary phase method tells us that the asymptotic $t\to\infty$ behavior of the solution $\phi$($r$, $t$, $\chi$) is given by singular points of
\[
{r^{1-\frac{N}{2}}}\sum_e E_e(\chi)\int_0\sp\infty{B_o}(r k)F(e, k, \omega )d k
\]
as a function of $\omega$, i. e. the singular point (\ref{10}). Therefore the study of analytic properties of the Green's function (\ref{25}) play a key role in understanding the late-time evolution of the wave-packet. The possible sources of non-analyticity in the Green's function are listed below,
\begin{itemize}
\item Branch points in the argument of the square root in $\omega$($k$, $r$, $\lambda$).
\item Pole at $\lambda$=0 coming from the defining integral.
\end{itemize}
The problem with branches of the argument of the square root in $\omega$($k$, $r$, $\lambda$) is absent in the Green's function $G$ when we approximate the double integration by an integration on a line defined by the relation of energy and momentum of a free particle, that is to say, by an integration on $\lambda\approx m\sp2$+$k\sp2$.

Since the $r_2$, $t_2$, $\chi_2$ space-time is important for the late-time evolution of the wave-packet, it is instructive to take a closer look at the approximation to the Green's function (\ref{25}) there. We proceed to apply this into the Green's function (\ref{25}), noticing that the behavior of $Z_i$ is fundamentally different depending on the sign of $\lambda$. I $f\lambda<$ 0, the imaginary exponent $it\omega$($k$, $r$, $\lambda$) in (\ref{23}) and (\ref{24}) is large and $Z_i$ goes to zero for all $r$, $t$, $\chi$. If $\lambda>$ 0, the real part of $it\omega$($k$, $r$, $\lambda$) is negative and hence $Z_i$ is finite for $r$, $t$, $\chi$. Applying this into the Green's function and using the asymptotic behavior of $Z_1$, $Z_2$ near infinity, we obtain,
\[
G(r_1, t_1, \chi_1; r_2, t_2, \chi_2)=
\]
\begin{equation}
={(r_1r_2)^{1-\frac{N}{2}}}\sum_e E_e(\chi_1)E_e\sp*(\chi_2)\int_0\sp\infty\frac{k^{2o}}{k\sp2+m\sp2}{B_o}(r k_1){B_o}(r k_2)d k.
\end{equation}
The case $\lambda>$ 0 is precisely where the inverse transformation should be in.
\subsection{From field to particle to force}
\label{Subseccion 5.6}
In field theory we define
\begin{equation}
\label{27}
W(J)=-\frac12\int\int d\sp4x d\sp4y J(x)G(x; y)J(y),
\end{equation}
where $J$($x$) is the so-called source function. We can choose any $J$($x$) we want and by exploiting this freedom of choice, we can extract a remarkable amount of physics.

We can go on to consider some possibilities for $J$($x$) (which we will refer to generically as sources), for example,
\[
J(x)=J_1(x)+J_2(x),\hbox{ where }J_a(x)=\delta\sp{(4)}(x, x_a).
\]
In other words, $J$($x$) is a sum of sources that are time-dependent infinitely sharp spikes located at $x_1$ and $x_2$ in space-time. (If the reader likes more mathematical rigor than is offered here, he is welcome to replace the delta function by lumpy functions peaking at $x_a$. The reader would simply clutter up the formulas without gaining much.) More picturesquely, we are describing two massive lumps sitting at $x_1$ and $x_2$ on the space-time, which are moving [time dependence in $J$($x$)].

What do the quantum fluctuations in the field $\phi$, that is, fluctuations in space-time, do to the lumps? If the reader expected an attraction between the two lumps, he is quite right.

$W$($J$) contains four terms. We neglect the ``self-interaction" term $J_1J_1$ since this contribution would be present in $W$ regardless of whether $J_2$ is present or not. We want to study the interaction between the two ``massive lumps" represented by $J_1$ and $J_2$. Similarly we neglect $J_2J_2$.

Plugging into (\ref{27}) and doing the integration over $d\sp4x$ and $d\sp4y$ we immediately obtain
\[
W(J)=-\frac1{4\pi\sqrt{r_1r_2}}\sum_{l=0}\sp\infty(1+2l)P_l(\chi_1\cdot\chi_2)\times
\]
\begin{equation}
\label{28}
\int_{-\infty}\sp\infty\int_0\sp\infty k\sp{1+2 l}\frac{e\sp{{i} (t_1 \omega(k, r_1, \lambda)-t_2 \omega(k, r_2, \lambda))}}\lambda{B_{\frac{1}{2}+l}}(r k_1){B_{\frac{1}{2}+l}}(r k_2)d k d\lambda.
\end{equation}
(The factor 2 comes from the two terms $J_2J_1$ and $J_1J_2$.)

Recall that in path integration formalism $Z$=${\mathcal C}e\sp{i W(J)}$ represents $e\sp{i E}$, where $E$ is the energy due to the presence of the two sources acting on each other. Setting $i W$=$i E$ we obtain from (\ref{28})
\[
E=-\frac1{4\pi\sqrt{r_1r_2}}\sum_{l=0}\sp\infty(1+2l)P_l(\chi_1\cdot\chi_2)\times
\]
\begin{equation}
\int_{-\infty}\sp\infty\int_0\sp\infty k\sp{1+2 l}\frac{e\sp{{i} (t_1 \omega(k, r_1, \lambda)-t_2 \omega(k, r_2, \lambda))}}\lambda{B_{\frac{1}{2}+l}}(r k_1){B_{\frac{1}{2}+l}}(r k_2)d k d\lambda.
\end{equation}

We will see that this energy is negative. The presence of two delta function sources, at $x_1$ and $x_2$, has lowered the energy. In other words, the two sources attract each other by virtue of their coupling to the field $\phi$. We have derived a first physical result in quantum field theory!

We identify $E$ as the potential energy between two sources. Even without doing the integrations we see that as the separation $x_1$, $x_2$ between the two sources becomes large, the oscillatory exponential cuts off the integration. The characteristic distance is the inverse of the characteristic value of $k$, which is $m$. Thus, we expect the attraction between the two sources to decrease rapidly to zero over the distance 1/$m$.

The range of the attractive force generated by the field $\phi$ is determined inversely by the mass $m$ of the particle described by the field.

We consider the first term
\begin{equation}
\frac1{4\pi\sqrt{r_1r_2}}P_0(\chi_1\cdot\chi_2)\int_{-\infty}\sp\infty\int_0\sp\infty\frac{k e\sp{{i} (t_1 \omega(k, r_1, \lambda)-t_2 \omega(k, r_2, \lambda))}}\lambda{B_{\frac{1}{2}}}(r k_1){B_{\frac{1}{2}}}(r k_2)d k d\lambda.
\end{equation}
This term is a double integration on the energy spectrum and the momentum of the particle.

Far from horizon $\rho \approx$ 0, we can approximate the double integration by an integration on a line defined by the relation of energy and momentum of a free particle, that is to say, by an integration on $\lambda\approx m\sp2$+$k\sp2$,
\begin{equation}
\frac1{4\pi\sqrt{r_1r_2}}P_0(\chi_1\cdot\chi_2)\int_0\sp\infty(m\sp2+k\sp2)\sp{-1}k{B_{\frac{1}{2}}}(r k_1){B_{\frac{1}{2}}}(r k_2)d k.
\end{equation}
We calculate
\begin{equation}
-\frac1{4\pi}r^{-\frac{1}{2}}P_0(\chi\cdot\chi_2)\int_0\sp\infty(m\sp2+k\sp2)\sp{-1}k{B_\frac12}(r k)\lim_{r_2\to0}r_2^{-\frac{1}{2}}{B_\frac12}(r k_2)d k
\end{equation}
\begin{equation}
=-\frac1{4\pi}\int_0\sp\infty\frac{2 k \sin(r k)}{({k^2}+{m^2}) \pi r}d k.
\end{equation}
The integral gives
\begin{equation}
\label{29}
E=-\frac1{4\pi r}e\sp{-m r}.
\end{equation}
The result is as we expected: The potential drops off exponentially over the distance scale 1/$m$. Obviously, $d E$/$d r>$ 0: The two massive lumps sitting on the space-time can lower the energy by getting closer to each other.

What we have derived is one of the most celebrated results in twentieth-century physics. Yukawa proposed that the attraction between nucleons in the atomic nucleus is due to their coupling to a field like the $\phi$ field described here. The known range of the nuclear force enabled him to predict not only the existence of the particle associated with this field, now called the $\pi$ meson\footnote{The etymology behind this word is quite interesting (A. Zea, {\it Fearful Symmetry}: see pp. 169 and 335 to learn, among other things, the French objection and the connection between meson and illusion)} or the peon, but its mass as well. As the reader probably knows, the peon was eventually discovered with essentially the properties predicted by Yukawa.

In deriving (\ref{29}) we direct our attention to the region far from the horizon. In this case, $\omega$ does not depend on $r$, i.e. $\omega$ is reduced to the ordinary relation in the case of a flat space-time. This means that our result (\ref{29}) is correct even if gravity  is taken into account.
\bibliographystyle{my-h-elsevier}

\end{document}